\documentclass[twocolumn]{jpsj3}
\usepackage{txfonts}
\usepackage{graphicx}% Include figure files
\usepackage{bm}% bold math
\usepackage[usenames]{color} 
\usepackage{amssymb} 
\usepackage{amsmath} 
\usepackage[utf8]{inputenc} 
\usepackage{braket}
\usepackage{mathptmx}

\newcommand{\bk}{\bm{k}}
\newcommand{\bq}{\bm{q}}

\title{Meissner Effect of Dirac Electrons in Superconducting State\\ due to Inter-band Effect}

\author{Tomonari Mizoguchi\thanks{mizoguchi@hosi.phys.s.u-tokyo.ac.jp} and Masao Ogata}
\inst{Department of Physics, The University of Tokyo, 7-3-1 Hongo, Bunkyo-ku, Tokyo 113-0033, Japan} %\\

\abst{Dirac electrons in solids show characteristic physical properties due to their linear dispersion relation and two-band nature. 
Although the transport phenomena of Dirac electrons in a normal state have intensively been studied, 
the transport phenomena in a superconducting state have not been fully understood.  
In particular, it is not clear whether Dirac electrons in a superconducting state show Meissner effect (ME), 
since a diamagnetic term of a current operator is absent as a result of the linear dispersion. 
We investigate the ME of three dimensional massive Dirac electrons in a superconducting state on the basis of Kubo formula, and clarify that Meissner kernel becomes finite by use of the inter-band contribution. 
This mechanism of the ME for Dirac electrons is completely different from that for the electrons in usual metals.
Our result shows that 
the Meissner kernel remains finite 
even when the superconducting gap vanishes.
This is an unavoidable problem in the Dirac electron system 
as reported in the previous works.
Thus, we use a prescription 
in which 
we subtract the normal state contribution.
In order to justify this prescription,
we develop 
a specific model where 
the Meissner kernel is obtained 
by the prescription.
We also derive the result for the electron gas by taking the non-relativistic limit of Dirac Hamiltonian, and clarify that the diamagnetic term of the Meissner kernel can be regarded as the inter-band contribution between electrons and positrons in terms of the Dirac model.}

\begin{document}
\maketitle

\section{Introduction}
Dirac electron systems have been attracting a great interest in recent years. 
One of the characteristic features of the Dirac electron systems is that its motion is described by Dirac equation instead of Schr\"odinger equation. 
Although the original Dirac equation describes the motion of relativistic (i.e., high-energy) electrons~\cite{dirac}, 
it is known that the electronic states of several materials with linear dispersion relation near the Fermi energy can be described by a low-energy effective model which has the same structure as Dirac equation. 
Examples of such materials are graphene~\cite{graphene_review}, $\alpha$-(BEDT-TTF)$_2$I$_3$~\cite{organic_review}, Bi~\cite{Wolff}, Ca$_3$PbO~\cite{Kariyado}, iron pnictide~\cite{feas,MT}, Na$_3$Bi~\cite{na3bi}, Cd$_3$As$_2$~\cite{cd3as2}, and so on. 
It has been reported that the physical properties of these materials are qualitatively different from those in the usual metals which are described by Schr\"odinger equation and Fermi liquid theory. 
Hence, it is important to investigate the fundamental properties of the Dirac electrons.\par

Up to now, many works have been carried out to investigate the normal-state properties of Dirac electron systems, such as Hall conductivity~\cite{fuseyaogata_Hall}, magnetoresistance~\cite{abrikosov,magres1}, orbital magnetism~\cite{fukuyamakubo,koshinoando,kobayashi}, 
Nernst coefficeint~\cite{igor},
and spin Hall conductivity~\cite{KM,fuseyaogata}. 
In these studies, it has been pointed out that the inter-band effect plays important roles, since the conduction and valence bands are close to each other. 
For instance, the giant orbital magnetism in Bi is due to the inter-band effect, and can not be explained by Landau-Periels's theory~\cite{La,Pe} of diamagnetism in which the inter-band is neglected.\par
On the other hand, the transport properties of Dirac electrons in a superconducting state have not been understood completely. 
In particular, it is not trivial whether the Dirac electrons in the superconducting state show the Meissner effect (ME), 
since the current operator does not have the diamagnetic term~\cite{remark}; 
for usual metals with a parabolic dispersion relation (i.e., in electron gas), the ME arises from the diamagnetic term of the current operator. 
Note that the diamagnetic current term appears when the Hamiltonian has a kinetic energy proportional to $\bm{k}^2$ with $\bm{k}$ being the momentum.~\cite{schrieffer}. 
It is therefore necessary to discuss the ME of Dirac electron in a superconducting state on the basis of the treatment which correctly includes the inter-band effect.\par
Although there have been many works on 
the superconductivity in the Dirac electron systems,
there are only a few studies on the ME~\cite{uchoa,kopnin}. 
In Ref. 23, the coexistence of the nodal charge density wave (CDW) 
and the superconductivity is studied.
Since the excitation spectrum of the mean-field CDW state has Dirac-like dispersion near the Fermi level,
the situation can be regarded as a superconductivity of a Dirac electron system.
However, their result shows that the Meissner kernel at zero temperature vanishes
when the Fermi level is on the Dirac point
although the superconducting order parameter 
$\Delta$ is nonzero.
This is rather unphysical.
As we show in the present paper,
the inter-band contribution 
gives a finite Meissner kernel
even when the density of states at the Fermi level is zero.
This indicates that the inter-band contribution is not taken into account 
correctly in Ref. 23.\par
In Ref. 24, the Meissner kernel of the superconducting state 
of graphene is calculated.
However,  the obtained result shows 
that the ME remains even in the normal state.
This is apparently unphysical. 
The authors of Ref. 24 suggests that this problem can be resolved
by subtracting the normal state contribution of the Meissner kernel.
However, they do not 
obtain the Meissner kernel
which remains after subtraction.\par
In this paper, we discuss the ME of a $4 \times 4$ massive Dirac electron system in the three-dimensional space assuming a s-wave Cooper pairing.
 The derivation of the Meissner kernel is based on Kubo formula. 
 We reveal that the inter-band effect plays an important role and that the ME appears as an inter-band contribution in spite of the absence of the diamagnetic term of the current operator.
Our results also show that the ME remains in the normal state as in Ref. 24.
This problem will be unavoidable as far as 
we use the unbounded Dirac dispersion.
Therefore, we use the prescription
to subtract the normal state contribution of the Meissner kernel
as in Ref. 24.
We discuss the obtained Meissner kernel
as a function of chemical potential.
In order to justify this prescription,
we develop
in Appendix 
a model
which is an extension 
of the Dirac Hamiltonian.
In this specific model,
we show that 
the Meissner kernel vanishes in the normal state,
and that the correct Meissner kernel
in the superconducting state
is obtained by the prescription.
Although we use a specific model,
we expect that 
this prescription is reasonable.\par
As discussed above, the inter-band effects 
in Dirac electron systems have been studied in the normal state.
On the other hand, 
it has not been recognized so much
that the inter-band effects play
important roles
in the response functions in the superconducting state.
In this paper, 
we show that
the inter-band effect is essential
to the Meissner kernel,
on the basis of the Kubo formula.\par
This paper is organized as follows. 
In the next section, we introduce a $4 \times 4$ massive Dirac Hamiltonian in the three-dimensional space, and the definition of the current operator in that model. 
We also show the treatment of the superconducting order parameter in the mean field approximation. 
In Sect. 3, 
we show the explicit form of the Meissner kernel in the present model
by use of Kubo formula,
and give the result of the numerical analysis. 
Then we discuss how Dirac electrons become the Meissner state. 
It will be shown that the inter-band contribution plays an important role in obtaining the finite Meissner kernel. 
We also mention the relation between the Dirac electron and the non-relativistic electron gas 
by considering the large band gap limit case.
We will show that our theory can reproduce the well-known results in the non-relativistic electron gas,
and that the origin of what we call the paramagnetic and diamagnetic terms originates from the intra and inter band term, respectively.
%We compare our result with the result of other Dirac electron system in Sect. 4. We point out that the ME due to the inter-band effect is ubiquitous in Dirac electron systems.
Finally, the brief summary is given in Sect. 4.
In Appendix, we develop a specific model 
in which the normal state Meissner kernel vanishes.
Part of the present work has been published before~\cite{mizoguchi}. 
\section{Formulation}
\subsection{Hamiltonian}
We consider the following $4\times4$ massive Dirac Hamiltonian in the three-dimensional space~\cite{fuseyaogata_Hall}:
\begin{equation}
H_0(\bk)=\hat{c_{\bk}^{\dagger}} \left(
\begin{array}{cc}
M \hat{I} & i v \bk \cdot \bm{\sigma} \\
-i  v \bk \cdot \bm{\sigma} & -M\hat{I} \\
\end{array}
\right)
\hat{c_{\bk}}, \label{diraceq}
\end{equation}
where $\hat{I}$ is the $2\times2$ unit matrix and $\bm{\sigma}=(\sigma_x,\sigma_y,\sigma_z)$ are Pauli matrices. 
$M$ is the band gap at the expanding center in the Brillouin zone, and $v$ is the Fermi velocity. 
The basis used in (\ref{diraceq}) is $\hat{c_{\bk}}= (c_{\bk,1,\uparrow},c_{\bk,1,\downarrow},c_{\bk,2,\uparrow},c_{\bk,2,\downarrow})^{\mathrm{T}}$ where $1,2$ denote the orbital, and $\uparrow,\downarrow$ denote labels of the time reversal pair.\par
\begin{figure}[t]
\begin{center}
\includegraphics[width=3.5cm]{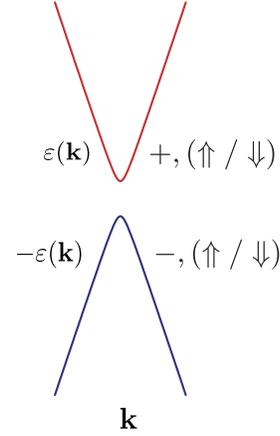}
\caption{(Color online) The dispersion relation of the $4\times4$ Dirac Hamiltonian in normal state. Red line corresponds to the upper band with index $+$, and blue line corresponds to the lower band with index $-$. Each band has two-fold degeneracy labelled by $\Uparrow$ and $\Downarrow$.} \label{hamiltonian}
\end{center}
\end{figure}

Diagonalizing this Hamiltonian, we obtain the following Hamiltonian,
\begin{align}
H_0(\bk)=\hat{a^{\dagger}_{\bk}}
\left(
\begin{array}{cc}
\varepsilon(\bk)\hat{I}&0 \\
0 & -\varepsilon(\bk)\hat{I} \\
\end{array}
\right)
\hat{a_{\bk}},
\end{align}
where $\varepsilon({\bm{k}})=\sqrt{M^2 + (vk)^2}$,
and $\hat{a_{\bk}}$ represents the band-basis which is expressed as $\hat{a_{\bk}}= (a_{\bk,+,\Uparrow},a_{\bk,+,\Downarrow},a_{\bk,-,\Uparrow},a_{\bk,-,\Downarrow})^{\mathrm{T}}$. 
Here, the indices $+$ and $-$ denote the upper and the lower bands respectively, and $\Uparrow$ and $\Downarrow$ denote psudo-spins corresponding to the two-fold degeneracy of each band [See Fig. \ref{hamiltonian}]. 
The unitary matrix which is defined as $\hat{a_{\bk}} = U(\bk) \hat{c_{\bk}} $ is given by
\begin{equation}
U(\bk) =
\left(
\begin{array}{cc}
X(\bm{k}) \hat{I}&- i  \bm{Y}(\bk) \cdot \bm{\sigma} \\
 - i  \bm{Y}(\bk) \cdot \bm{\sigma}  & X(\bk) \hat{I} \\
\end{array}
\right), \label{unitary_matrix}
\end{equation}
where $X(\bk)=\sqrt{\frac{\varepsilon(\bk)+M}{2\varepsilon(\bk)}}$ and $\bm{Y}(\bk) = \sqrt{\frac{1}{2\varepsilon(\bk)(\varepsilon(\bk)+M)}}v\bk$.

\subsection{Current operator}
In this subsection, we discuss the current operator in the absence and presence of the vector potential. Without the vector potential, $\bm{A}$, the current operator in the momentum space is given by 
\begin{align}
\hat {\bm{j}} (\bm{q}) = & -e \sum_{\bm{k}} \hat{c}^{\dagger}_{\bm{k}-\bm{q}} \partial_{\bm{k}} H_0(\bm{k}) \hat{c}_{\bm{k}} \notag \\
 =  & -e \sum_{\bm{k}} \hat{c}_{\bm{k}-\bm{q}}^{\dagger}
\left(
\begin{array}{cc}
0 & \mathrm{i} v \bm{\sigma} \\
- \mathrm{i} v \bm{\sigma} & 0 \\
\end{array}
\right) \hat{c}_{\bm{k}}\notag \\
\equiv &   \sum_{\bm{k}} \hat{c}_{\bm{k}-\bm{q}}^{\dagger}\bm{j} \hat{c}_{\bm{k}}, \label{current}
\end{align}
where $e>0$ is the absolute value of the charge of an electron. 
The vector potential $\bm{A}$ is introduced in the Hamiltonian by replacing $\bk$ by $\bk+e\bm{A}$. 
However, we can readily see that the current operator in Eq. (\ref{current}) does not change even in the presence of $\bm{A}$, since $H_{0}(\bk)$ contains only linear terms with respect to $\bk$. \par
This property of the current operator is essentially different from that in the electron gas
in which the current operator is given by $\hat{\bm{j}} (\bq)= \sum_{\bk,\sigma} c^{\dagger}_{\bk-\bq,\sigma} \frac{e}{m}(\bk +e\bm{A} )c_{\bk,\sigma}$ where the last term is called as the diamagnetic term. 
Apparently, the current operator of the Dirac electron systems does not have diamagnetic current.
As we will show shortly, the absence of the diamagnetic term is crucial for discussing the mechanism of the ME in Dirac electron in superconducting state.
%---------%
%---------%

\subsection{Superconductivity}
In this subsection, we introduce superconductivity in the Dirac electron system. 
We assume the following attractive interaction,
\begin{equation}
 H_{\mathrm{int}} = -V\sum_{|\bk| ,|\bk^{\prime}| < k_c} \sum_{\eta,\eta^{\prime} = \pm} a^{\dagger}_{\bk,\eta,\Uparrow} a^{\dagger}_{-\bk,\eta,\Downarrow} a_{-\bk^{\prime},\eta^{\prime},\Downarrow} a_{\bk^{\prime},\eta^{\prime},\Uparrow},
 \end{equation}
where $k_c$ is the cutoff momentum corresponding to the range of energy in which the attractive interaction works. 
When $vk_c \gg M$,
a superconducting state is realized even if the chemical potential is located in the band gap
(i.e., there is no density of states at the Fermi level) for sufficiently large $V$, since electrons with $k<k_c$ can still contribute to the Cooper pairing. \par
\begin{figure}[t]
\begin{center}
\includegraphics[width=6cm]{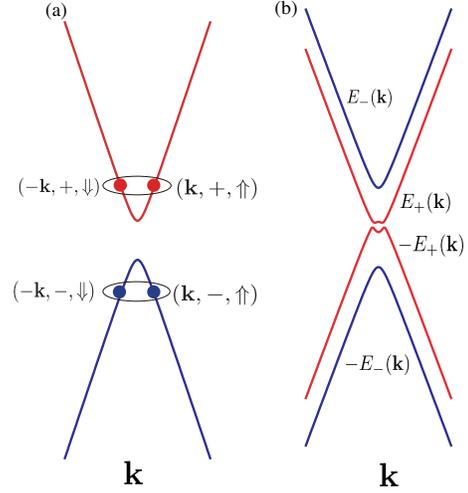}
\caption{(Color online) (a) Schematic picture of the cooper pairing, and (b) energy eigenvalues for Bogoliubov quasi-particles.} \label{cooper}
\end{center}
\end{figure}

We define the superconducting order parameter in the s-wave symmetry (see Fig. \ref{cooper}(a)),
\begin{equation}
\Delta_{\pm}= V \sum_{| \bk^{\prime}| < k_c} \langle a_{-\bk^{\prime},\pm,\Downarrow} a_{\bk^{\prime},\pm,\Uparrow} \rangle.
\end{equation}
By applying the mean field approximation to the Hamiltonian, $H=H_{0}+H_{\mathrm{int}}$, we obtain the BCS Hamiltonian,
\begin{align}
H_{\mathrm{BCS}} = \sum_{\bk} \sum_{\eta=\pm} \sum_{\Sigma=\Uparrow,\Downarrow} \xi_{\pm}(\bk)a^{\dagger}_{\bk,\eta,\Sigma} a_{\bk,\eta,\Sigma} \notag \\
-\sum_{\bk} \sum_{\eta=\pm}\Delta(\bk)[a^{\dagger}_{\bk,\eta,\Uparrow}a^{\dagger}_{-\bk,\eta,\Downarrow} + a_{-\bk,\eta,\Downarrow}a_{\bk,\eta,\Uparrow}] \label{bcs}
\end{align}

where $\Delta(\bk)=(\Delta_{+} + \Delta_-)\Theta (k_c-|\bk|) = \Delta\Theta (k_c-|\bk|)$ with $\Theta(x)$ being the step function,
 and $\xi_{\pm}(\bk)=\pm \varepsilon(\bk) -\mu$.\par 
Diagonalizing $H_{\mathrm{BCS}}$, we obtain the the excitation energy $E_{\pm}(\bk) \equiv \sqrt{\xi_\pm(\bk)^2+\Delta^2(\bk)}$ [see Fig. \ref{cooper}(b)].
Then, thermal Green's function, 
\begin{equation}
[\mathcal{G}(\bk,i \omega_n)]_{\alpha_1\alpha_2} \equiv - \int_{0}^{\beta} d\tau e^{i \omega_n\tau } \langle T_{\tau}[a_{\bk,\alpha_1}(\tau)a^{\dagger}_{\bk,\alpha_2}(0)]\rangle,
\end{equation}
 and the anomalous Green's functions, 
 \begin{subequations}
 \begin{equation}
 [\mathcal{F}(\bk,i \omega_n)]_{\alpha_1\alpha_2} \equiv - \int_{0}^{\beta} d\tau \ e^{i \omega_n\tau } \langle T_{\tau}[a_{\bk,\alpha_1}(\tau)a_{-\bk,\alpha_2}(0)] \rangle,
 \end{equation} 
  \begin{equation}
[\mathcal{F}^{\dagger}(\bk ,i \omega_n)]_{\alpha_1\alpha_2} \equiv -\int _0^{\beta}  d\tau e^{i \omega_n \tau}\langle T_{\tau}[a^{\dagger}_{-\bk,\alpha_1}(\tau)a^{\dagger}_{\bk,\alpha_2}(0)] \rangle,
  \end{equation}
   \end{subequations}
 are obtained in the form of $4\times4$ matrices as follows. 
(Note that $\alpha_1$ and $\alpha_2$ are the sets of band and pseudo spin indices.)
\begin{equation}
\mathcal{G}(\bk,i \omega_n)=
\left(
\begin{array}{cc}
\mathcal{G}_+(\bk ,i \omega_n)  \hat {I}& 0 \\
0& \mathcal{G}_-(\bk ,i \omega_n) \hat {I}  \\
\end{array}
\right),
\end{equation}
and 
\begin{align}
\mathcal{F}(\bk,i \omega_n)
= &-\mathcal{F}^{\dagger}(\bk,i \omega_n) \notag \\
= & 
\left(
\begin{array}{cc}
 i \sigma_y \mathcal{F}_+(\bk ,i \omega_n) &0   \\
0 & i \sigma_y \mathcal{F}_-(\bk ,i \omega_n)  \\
\end{array}
\right),
\end{align}
where
\begin{equation} 
\mathcal{G}_{\pm}(\bk,i \omega_n) = \frac{u^2_{\pm}(\bk)}{i\omega_n-E_{\pm}(\bk)} + \frac{v^2_{\pm}(\bk)}{i \omega_n+E_{\pm}(\bk)}, 
\end{equation}
and
\begin{align} 
\mathcal{F}_{\pm}(\bk,i  \omega_n) = -u_{\pm}(\bk)v_{\pm}(\bk)(\frac{1}{i \omega_n-E_{\pm}(\bk)}-\frac{1}{i \omega_n+E_{\pm,}(\bk)}), 
\end{align}
with $u^{2}_{\pm}(\bk)=\frac{1}{2}[1+\frac{\xi_{\pm}(\bk)}{E_{\pm}(\bk)}]$, $v^{2}_{\pm}(\bk)=\frac{1}{2}[1-\frac{\xi_{\pm}(\bk)}{E_{\pm}(\bk)}]$, 
and $u_{\pm}(\bk)v_{\pm}(\bk) = \frac{\Delta(\bk)}{2E_{\pm}(\bk)}$. 
It should be noted that the upper and lower parts of these Green's functions are decoupled with each other. 

%---------%
%---------%

\section{Meissner effect in Dirac electron systems}
\subsection{Small band gap case: Dirac electrons in solids}
Using the formulations derived in the previous section, we calculate Meissner kernel, $K^{\mathrm(S)}_{xx}$, through Kubo formula. 
Since there is no diamagnetic term of the current operator, $K^{\mathrm(S)}_{xx}(\textbf{q},i \omega_{\lambda})$ is given only by the current-current correlation function,
\begin{equation}
K^{\mathrm(S)}_{xx}(\bq,i \omega_{\lambda}) =- \int_{0}^{\beta} d \tau e^{i\omega_{\lambda} \tau} \langle T_\tau (\hat{j}_x(\bq,\tau) \hat{j}_x(-\bq,0))\rangle, \label{ker1}
\end{equation}
where $\omega_{\lambda} =\frac{2\pi}{\beta}\lambda \ (\lambda = 0,\pm1,\pm2, \cdots)$ denotes a bosonic Matubara frequency and $\tau$ an imaginary time. 
The Meissner kernel for real frequency is given by the analytic continuation $K^{\mathrm(S)}_{xx}(\textbf{q},\omega)=K^{\mathrm(S)}_{xx}(\textbf{q},i \omega_{\lambda}) |_{i \omega_{\lambda} \rightarrow \omega }$.\par
Applying the Bloch-De Dominics theorem to Eq. (\ref{ker1}), we obtain the following expression of $K_{xx}(\bq,i \omega_\lambda)$: 
\begin{align}
 & K^{\mathrm(S)}_{xx}(\bm{q},i \omega_{\lambda}) \notag \\ 
= &  T \sum_{\bm{k},\omega_n}\mathrm{Tr}
  [\mathcal{G}(\bm{k}-\bm{q},\omega_n-\omega_{\lambda})\tilde{j}_x(\bk,\bq) 
\mathcal{G}(\bm{k},\omega_n) \tilde{j}_x(\bk-\bq,-\bq)  \notag \\
-  & \mathcal{F}^{\dagger}(\bm{k}-\bm{q},\omega_n-\omega_{\lambda})\tilde{j}_x(\bk,\bq) 
\mathcal{F}(\bm{k},\omega_n) \tilde{j}^{\mathrm{T}}_x(-\bk,-\bq)], 
\label{kernel_kubo}
\end{align}
where $\tilde{j}_x(\bk,\bq)$ is given by $\tilde{j}_x(\bk,\bq) \equiv [U(\bm{k}-\bm{q}) j_xU^{\dagger}(\bm{k})]$, 
and $\omega_{n} =\frac{(2n+1)\pi}{\beta} (n = 0,\pm1,\pm2, \cdots)$ denotes a fermionic Matubara frequency.
It should be noted that the limit $\omega\rightarrow 0$ has to be taken before the limit $\bq \rightarrow 0$.\par

By taking the trace in Eq. (\ref{kernel_kubo}), we obtain the Meissner kernel which consists of two parts as 
\begin{equation}
K^{\mathrm(S)}_{xx}(\bm{q},0) = K^{\mathrm{intra,(S)}}_{xx}  (\bm{q},0) + K^{\mathrm{inter},(S)} _{xx}(\bm{q},0).
\end{equation}
$K^{\mathrm{intra,(S)}}_{xx}  (\bm{q},0)$ comes from the term in which the two Green's functions in Eq. (\ref{kernel_kubo}), ($\mathcal{G}\mathcal{G}$ or $\mathcal{F}^{\dagger}\mathcal{F}$) have the same band indeces $(+,+)$ or $(-,-)$, 
and $K^{\mathrm{inter,(S)}}_{xx}  (\bm{q},0)$ from the Green's functions with the opposite band indeces, i.e., $(+,-)$ or $(-,+)$.

%---------%
\begin{figure*}[t]
\begin{center}
\includegraphics[width=17cm]{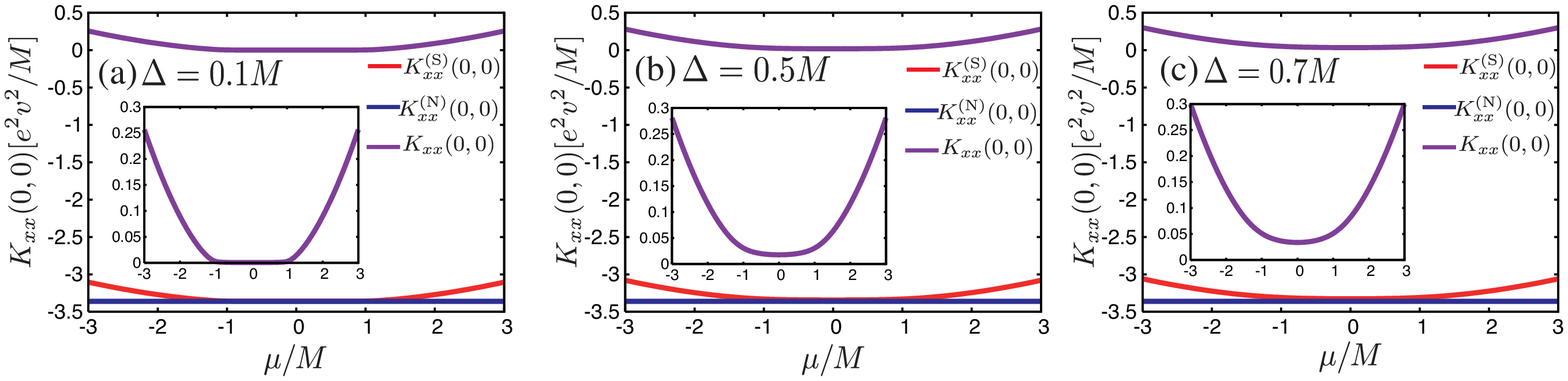}
\caption{(Color online) $\mu$ dependence of $K^{\mathrm{(S)}}_{xx}(0,0)$, $K^{\mathrm{(N)}}_{xx}(0,0)$, and $K_{xx}(0,0)$ for (a) $\Delta=0.1M$, (b) $\Delta=0.1M$, and (c) $\Delta(\bk)=0.7M$. $\mu$ dependence of $K_{xx}$ is in the inset.} \label{kernel}
\end{center}
\end{figure*}
%---------%

The explicit forms of these are obtained as 
\begin{align}
K^{\mathrm{intra,(S)}}_{xx}  (\bm{q},0) = - \frac{e^2 v^2}{2} \sum_{\bm{k}} \sum_{\eta=\pm} g_{xx}^{\mathrm{intra}}(\bm{k},\bm{q}) \notag \\
\times [B_\eta^{\mathrm{intra(i)}} (\bk,\bq) \frac{f(E_\eta(\bm{k}))-f(E_\eta\bm{k}-\bm{q}))}{E_\eta(\bm{k}-\bm{q})-E_\eta (\bm{k})}  \notag\\
- B_\eta^{\mathrm{intra(ii)}} (\bk,\bq)\frac{f(E_\eta(\bm{k}))+f(E_\eta(\bm{k}-\bm{q}))-1}{E_\eta(\bm{k}-\bm{q})+E_\eta(\bm{k})} ], 
\label{intrafiniteq}
\end{align}
and
 \begin{align}
 K^{\mathrm{inter,(S)}}_{xx}  (\bm{q},0) = \frac{e^2 v^2}{2} \sum_{\bm{k}} \sum_{\eta=\pm} g_{xx}^{\mathrm{inter}}(\bm{k},\bm{q}) \notag \\
 \times [B^{\mathrm{inter(i)}}(\bk,\bq)\frac{f(E_{-\eta}(\bm{k}))-f(E_{\eta}(\bm{k}-\bm{q}))}{E_{\eta}(\bm{k}-\bm{q})-E_{-\eta}(\bm{k})} \notag \\
-B^{\mathrm{inter(ii)}}(\bk,\bq)\frac{f(E_{-\eta}(\bm{k}))+f(E_{\eta}(\bm{k}-\bm{q}))-1}{E_{\eta}(\bm{k}-\bm{q})+E_{-\eta}(\bm{k})} ]. 
\label{interfiniteq}
 \end{align}
 Here, $f(x)$ is the Fermi distribution fucntion, and $g_{xx}^{\mathrm{intra}} (\bm{k},\bm{q}) $, $g_{xx}^{\mathrm{inter}} (\bm{k},\bm{q})$ are given by
 \begin{subequations}
\begin{gather}
g_{xx}^{\mathrm{intra}} (\bm{k},\bm{q}) =  \frac{v ^2 [k_x(k_x-q_x)- k_y(k_y-q_y) - k_z(k_z-q_z)]}{\epsilon(\bm{k})\epsilon(\bm{k}-\bm{q})} \notag \\
+ \frac{\epsilon(\bm{k})\epsilon(\bm{k} - \bm{q} )-M^2}{\epsilon(\bm{k})\epsilon(\bm{k}-\bm{q})},
\end{gather}
\begin{gather}
g_{xx}^{\mathrm{inter}} (\bm{k},\bm{q}) = \frac{v^2[ k_x(k_x-q_x)- k_y(k_y-q_y) - k_z(k_z-q_z)]}{\epsilon(\bm{k})\epsilon(\bm{k}-\bm{q})} \notag \\
-\frac{\epsilon(\bm{k})\epsilon(\bm{k}-\bm{q})+M^2 }{\epsilon(\bm{k})\epsilon(\bm{k}-\bm{q})}.
\end{gather}
\end{subequations}
In Eqs. (\ref{intrafiniteq}) and (\ref{interfiniteq}), $B$'s are coherence factors,
\begin{subequations}
\begin{equation}
B^{\mathrm{intra(i)}}_{\eta}(\bm{k},\bm{q})=1+\frac{\xi_{\eta}(\bm{k}-\bm{q})\xi_{\eta}(\bm{k})+\Delta^2(\bk)}{E_{\eta}(\bm{k}-\bm{q})E_{\eta}(\bm{k})},
\end{equation}
\begin{equation}
B^{\mathrm{intra(ii)}}_{\eta}(\bm{k},\bm{q})=1-\frac{\xi_{\eta}(\bm{k}-\bm{q})\xi_{\eta}(\bm{k})+\Delta^2(\bk)}{E_{\eta}(\bm{k}-\bm{q})E_{\eta}(\bm{k})},
\end{equation}
\begin{equation}
B^{\mathrm{inter(i)}}(\bm{k},\bm{q})=1+\frac{\xi_+(\bm{k}-\bm{q})\xi_{-}(\bm{k})-\Delta^2(\bk)}{E_+(\bm{k}-\bm{q})E_{-}(\bm{k})},
\end{equation}
\begin{equation}
B^{\mathrm{inter(ii)}}(\bm{k},\bm{q})=1-\frac{\xi_+(\bm{k}-\bm{q})\xi_{-}(\bm{k})-\Delta^2(\bk)}{E_+(\bm{k}-\bm{q})E_{-}(\bm{k})}.
\end{equation}
\end{subequations}

When we take the limit of $\bq \rightarrow \bm{0}$, we obtain $K^{\mathrm{(S)}}_{xx}(0,0) = K^{\mathrm{intra,(S)}}_{xx}  (0,0) + K^{\mathrm{inter,(S)}} _{xx}(0,0)$ with 
\begin{align}
K^{\mathrm{intra,(S)}}_{xx} (0,0) =  2e^2 v^4\sum_{|\bm{k}|<k_c}\frac{k_x^2}{\varepsilon^2(\bk)} [\frac{\partial f(E_{+}(\bk))}{\partial E_+(\bk)} + \frac{\partial f(E_-(\bk))}{\partial E_-(\bk)}] ,\label{kintra}
\end{align}
and
 \begin{align}
 K^{\mathrm{inter,(S)}}_{xx}  (0,0) = - 2 e^2 v^2\sum_{\bm{k}}\left(1-\frac{v^2k_x^2}{\varepsilon^2(\bk)}
\right) \notag \\
  \times [\left(1+\frac{\xi_+(\bm{k})\xi_-(\bm{k})-\Delta^2(\bk)}{E_+(\bm{k})E_-(\bm{k})}\right) \frac{f(E_-(\bm{k}) )-f(E_+(\bm{k})) }{E_+(\bm{k})-E_-(\bm{k})} \notag \\
-\left(1-\frac{\xi_+(\bm{k})\xi_-(\bm{k})-\Delta^2(\bk)}{E_+(\bm{k})E_-(\bm{k})} \right) \frac{f(E_-(\bm{k}))+f(E_+(\bm{k}))-1}{E_+(\bm{k})+E_-(\bm{k})}] . \label{kinter}
 \end{align} 
 
It should be noted that for $T=0$, $K^{\mathrm{intra,(S)}}_{xx}(0,0)$ is 0 since the spectrum has a finite gap at the chemical potential. 
Hence, $K^{\mathrm{(S)}}_{xx}$ only consists of the inter-band contribution.
This is in sharp contrast to the conventional s-wave BCS case, in which the Meissner kernel from the paramagnetic current vanishes at $T=0$, but the Meissner kernel from the diamagnetic current remains which leads to the ME. In contrast, in the Dirac case, the diamagnetic current does not exist from the beginning, but instead the inter-band contribution exists, which leads to the ME.\par

However, there is an interesting problem appearing from the inter-band contribution. 
Even if we put $\Delta(\bk)=0$, i.e., in the normal state, the kernel $K^{\mathrm{(N)}}_{xx}(0,0)$ does not vanish.
The explicit form is 
\begin{align}
 K^{\mathrm{intra,(N)}}_{xx} (0,0) = &   2e^2 v^2 \sum_{\bk} \frac{v^2 k_x^2}{\varepsilon^2(\bk)} \left[\frac{\partial f(\xi_{+}(\bk))}{\partial \xi_+(\bk)} + \frac{\partial f(\xi_-(\bk))}{\partial \xi_-(\bk)} \right] \notag \\
 = & - \frac{e^2v^2}{3\pi^2} \frac{(\mu^2-M^2)^{3/2} \Theta(|\mu|-M)}{|\mu|v^3} ,
\end{align}
and
\begin{align}
& K^{\mathrm{inter,(N)}}_{xx} (0,0) \notag \\ = &  -  2e^2 v^2 \sum_{\bm{k}}\left(1-\frac{v^2k_x^2}{\varepsilon^2(\bk)}\right)
 \left[ \frac{f(\xi_-(\bm{k}))-f(\xi_+(\bm{k}) ) }{\varepsilon(\bk)}\right] \notag \\
 = & - \frac{e^2v^2}{3\pi^2}\left[ \frac{\Lambda^3}{\varepsilon(\Lambda)}- \frac{(\mu^2-M^2)^{3/2} \Theta(|\mu|-M)}{|\mu|v^3}\right],
\end{align}
hence 
\begin{align}
K^{\mathrm{(N)}}_{xx} (0,0)  &= K^{\mathrm{intra,(N)}}_{xx} (0,0)  + K^{\mathrm{inter,(N)}}_{xx} (0,0)  \notag \\
= &  - \frac{e^2v^2}{3\pi^2}\left[ \frac{\Lambda^3}{\varepsilon(\Lambda)} \right].
\end{align}
Here we introduce the ultraviolet cut-off momentum $\Lambda$.
This finite Meissner kernel in the normal state
is due to the linear dispersion and the unboundedness of the spectra in the Dirac electron systems. 
A similar problem was pointed out before in graphene~\cite{kopnin}.
In the usual electron gas model,
$K^{\mathrm{(N)}}_{xx}=0$ holds since the paramagnetic term exactly cancels with the diamagnetic term.
However, in the case of Dirac electron systems, 
there is no \lq \lq counter contribution"
by which the normal state Meissner kernel vanishes. 
Therefore, in the following, 
we use the prescription to subtract the $K^{\mathrm{(N)}}_{xx}$ 
as suggested in Ref. 24, 
i.e., we calculate the Meissner kernel in the superconducting state by
\begin{equation}
K_{xx}\equiv K^{\mathrm{(S)}}_{xx}-K^{\mathrm{(N)}}_{xx}. \label{true_kernel}
\end{equation}
It should be noted that 
only the contribution from $|\bk|<k_c$,
where $\Delta(\bk)$ is finite, 
gives the finite value in Eq. (\ref{true_kernel}),
since the contribution from $|\bk|>k_c$  
in the normal and superconducting states
are exactly same
and cancel with each other.
In order to justify the prescription Eq. (\ref{true_kernel}),
we develop 
in Appendix a model 
which is an extension of the Dirac Hamiltonian
and show that the Meissner kernel in the normal state vanishes. 
In this specific model,
we also show that the correct Meissner kernel 
in the superconducting state is 
obtained by the prescription
of (\ref{true_kernel}).
We should note that 
it is not clear whether this prescription can be applied to general Dirac electron systems,
although it is justified in the specific model in Appendix.
However, it is physically required that 
the Meissner kernel should vanishes 
in the normal state
and we expect that the prescription (\ref{true_kernel})
is reasonable.\par
 By using Eq. (\ref{true_kernel}), we numerically calculate $K_{xx}(0,0)$ at $T=0$. 
In the numerical calculation, we set $k_c=10M/v$.\par 
Figures \ref{kernel} shows $\mu$ dependence of $K^{\mathrm{(S)}}_{xx}, K^{\mathrm{(N)}}_{xx}$, 
and $K_{xx}$ for a fixed $\Delta$, (a) $\Delta=0.1M$, (b) $\Delta=0.5M$ , and (c) $\Delta=0.7M$ respectively.
(The insets show $K_{xx}$ in the extended scale.)
We can see that although $K_{xx}^{\mathrm{(S)}}$ and  $K_{xx}^{\mathrm{(N)}}$ are both negative, 
$K_{xx}(0,0)$ is always positive and finite, thus we obtain the Meissner state.\par
We can also see in Figs. \ref{kernel}, that $K_{xx}$ strongly depends on $\Delta$, in $-M \leq \mu \leq M$. 
This is another sharp contrast to the conventional BCS case in which $K_{xx}$ does not depend on $\Delta$ and depends only on the carrier density. 

\subsection{Large band gap case: non-relativistic electron}
%---------%
\begin{figure}[b]
\begin{center}
\includegraphics[width=3cm]{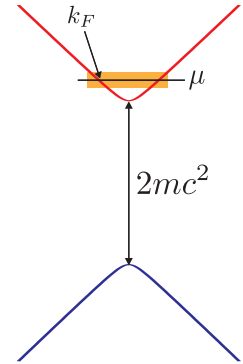}%---------%
\end{center}
\caption{(Color online) Dispersion relation in the non-relativistic limit. Yellow shade denotes the range where the attractive interaction works.}
 \label{nonrela}
\end{figure}
%---------%

Up to now, we consider the case in which the band gap is very small ($\sim 10$ meV). 
In this subsection, we consider the opposite case, i.e., the large band gap case.
In particular, it is important to test whether the present theory reproduces the results for non-relativistic theory.
To access to the non-relativistic limit, 
we replace $M$ to $mc^2$, and $v$ to $c$, 
with $m$ being the mass of an electron and $c$ the speed of light. 
Then, we can regard the upper band as the real electron, and the lower band as the positron (see Fig. \ref{nonrela}).
We further assume that the chemical potential is located in the upper band, 
and that $\mu-mc^2$ is much smaller than $mc^2$.
Then we obtain the Meissner kernel in the normal state as 
\begin{align}
K^{\mathrm{intra,(N)}}_{xx} (0,0) =- \frac{e^2c^2}{3\pi^2} \frac{(\mu^2-m^2c^4)^{3/2}}{\mu c^3} \sim -\frac{e^2 n}{m}, \label{lg1}
\end{align}
and 
\begin{align}
K^{\mathrm{inter,(N)}}_{xx} (0,0) 
 = & - \frac{e^2c^2}{3\pi^2} \left[ \frac{\Lambda^3}{\varepsilon(\Lambda)} - \frac{(\mu^2-m^2c^4)^{3/2}}{\mu c^3} \right] \notag \\
 \sim & \frac{e^2 n}{m} - \frac{e^2c^2}{3\pi^2} \frac{\Lambda^3}{\varepsilon(\Lambda)},
\label{lg2}
\end{align}
where $k_F=\sqrt{2m(\mu-mc^2)}$ is the Fermi momentum, 
and $n=\frac{k_F^3}{3\pi^2}$ is the electron density.
Here we use the approximantion $\mu - mc^2 \ll mc^2$. \par
Comparing Eqs. (\ref{lg1}) and (\ref{lg2}) with the results for the electron gas~\cite{schrieffer}, we can see that $K_{xx}^{\mathrm{intra},(N)}$ is exactly equal to the paramagnetic term in the electron gas 
and that $K_{xx}^{\mathrm{inter},(N)}$ is exactly equal to the diamagnetic term apart from the divergent part
($- \frac{e^2c^2}{3\pi^2} \frac{\Lambda^3}{\varepsilon(\Lambda)}$). 
Note that this divergent term comes from the region far away from the Fermi surface, 
and that this is exactly what we deal with by the subtraction in order to avoid the unphysical result in the last subsection.
We think that this kind of divergence is inevitable within the Dirac theory.
Eqs. (\ref{lg1}) and (\ref{lg2}) indicate, first, that we can reproduce the result of the electron gas starting from the Dirac model; 
second, that the diamagnetic term of the kernel in the electron gas originates from the inter-band contribution between positrons and electrons.
\par
Now we turn to the superconducting state.
We assume that 
the range of the energy in which the attractive interaction works is much smaller than $mc^2$ or $\mu-mc^2$ as shown in Fig. \ref{nonrela}.
Thus, the pairing occurs in the very vicinity of the Fermi level on the upper band [see the yellow shade in Fig. \ref{nonrela}].
In this case, $K_{xx}^{\mathrm{intra,(S)}}$ becomes 0 as we have discussed in the last subsection.
On the other hand, $K_{xx}^{\mathrm{inter,(S)}}$ is almost same as $K_{xx}^{\mathrm{inter,(N)}}$,
since the states below the Fermi level do not change by the pairing except for the vicinity of the Fermi surface.
Therefore, we obtain the Meissner kernel in the superconducting state as 
\begin{equation}
K_{xx}^{\mathrm{(S)}} \sim K_{xx}^{\mathrm{inter,(N)}} \sim \frac{e^2 n}{m},
\end{equation}
which is completely consistent with the result of the superconducting state in electron gas.

\section{Summary}
In summary, we have investigated the ME of three-dimensional massive Dirac electron in superconducting state on the basis of Kubo formula. 
Although the diamagnetic term of the current operator is absent in Dirac electron system, the Meissner kernel finite for any value of the chemical potential, since the inter-band contribution remains finite.
%We have to care about the finite Meissner kernel in the normal state, but this can be eliminated by subtracting the normal component.
This inter-band mechanism of the ME of Dirac electron is essentially different from that of usual metals.\par
In the Dirac electron system,
there is an unavoidable problem 
that the Meissner kernel 
remains finite 
even in the normal state.
In order to obtain the Meissner kernel in the superconducting state,
we use a prescription of Eq. (\ref{true_kernel}).
Although we have not proved that this prescription is completely justified,
we develop a specific model in which the Meissner kernel
vanishes in the normal state and the prescription is reasonable. \par
We also have studied the large band-gap limit to discuss the non-relativistic case.
We clarify the relation that the paramagnetic term of the kernel originates from the intra-band term and the diamagnetic term from the inter-band term.

\begin{acknowledgment}
TM is grateful to Y. Masaki for fruitful discussions and comments. 
This work was supported by Grants-in-Aid for Scientific Research (A) (No.24244053). 
\end{acknowledgment}

\appendix
\section{Meissner effect in Dirac electron systems with a quadratic term}
\begin{figure}[b]
\begin{center}
\includegraphics[width=5cm]{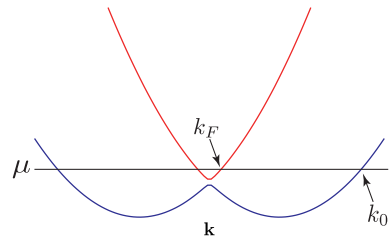}
\caption{(Color online) The dispersion relation of the $4\times4$ Dirac Hamiltonian in normal state with the quadratic term.} \label{dis_q}
\end{center}
\end{figure}

In this appendix, we perform a simple model calculation in which the Meissner kernel in the normal state vanishes.
In order to avoid the unboundedness of the Dirac dispersion, 
we add a kinetic energy term $\frac{a}{2}k^2$ in the Hamiltonian instead of introducing the Brillouin zone,
\begin{equation}
\tilde H_0(\bm{k})=\hat{c_{\bm{k}}^{\dagger}} \left(
\begin{array}{cc}
(M+\frac{a}{2}k^2) \hat{I} & \mathrm{i} v \bm{k} \cdot \bm{\sigma} \\
-\mathrm{i}  v \bm{k} \cdot \bm{\sigma} & (-M+\frac{a}{2}k^2) \hat{I} \\
\end{array}
\right)
\hat{c_{\bm{k}}}, 
\end{equation}
where parameter $a$ is assumed to be very small. 
The energy dispersion becomes $\varepsilon({\bm{k}})+\frac{a}{2}k^2$, and 
a new Fermi surface appears as shown in Fig. \ref{dis_q}. 
The matrix form of the current operator, Eq.\ (\ref{current}) changes as
\begin{align}
\hat {\bm{j}} (\bm{q})  
 =  -e \sum_{\bm{k}} \hat{c}_{\bm{k}-\bm{q}}^{\dagger}
\left(
\begin{array}{cc}
a (\bm{k} + e{\bm A})  \hat{I}  & iv \bm{\sigma} \\
-iv \bm{\sigma} & a (\bm{k} + e{\bm A}) \hat{I} \\
\end{array}
\right)
\hat{c}_{\bm{k}}. \label{NewCurrent}
\end{align}
Note that there appears a \lq\lq diamagnetic part" of the current, 
$-ea(\bk +e\bm{A}) \hat{I}$.
However, it is proportional to the small parameter $a$,
and we will show that the 
main part of the Meissner kernel in the superconducting state comes 
not from this \lq\lq diamagnetic part" 
but from the 
inter-band contribution as discussed in Sect. 3.

%---------%
%---------%

We can carry out a similar calculation of the Meissner kernel as in Sect. 3.
Note that the unitary matrix in Eq.\ (\ref{unitary_matrix}) does not change. 
After some algebra, we obtain
\begin{align}
K^{\mathrm{intra,(S)}}_{xx} (0,0) =  2e^2 \sum_{\bm{k},\eta=\pm} \frac{(v^2k_x+ \eta ak_x \varepsilon(\bm{k}))^2}{\varepsilon^2(\bm{k})} \frac{\partial f(E_{\eta}(\bm{k}))}{\partial E_\eta (\bm{k})} ,
\label{ArtifKintra}
\end{align}
and
\begin{align}
K^{\mathrm{inter,(S)}}_{xx} (0,0) &= -2 e^2 v^2 \sum_{\bm{k}}\left( 1-\frac{v^2k_x^2}{\varepsilon^2(\bm{k})} \right) \notag \\
&\times \frac{1}{E_+(\bm{k})+E_-(\bm{k})}
\left( 1-\frac{\xi_+(\bm{k})\xi_-(\bm{k})-\Delta^2(\bk)}{E_+(\bm{k})E_-(\bm{k})} \right) \notag \\
&\equiv  \sum_{\bm{k}} F^{\mathrm{inter}}(\bm{k},\Delta(\bm{k})),
\label{ArtifKinter}
\end{align} 
at $T=0$ where $E_{\pm}(\bm{k}) \equiv \sqrt{\xi_\pm(\bm{k})^2+\Delta(\bk) ^2}$, and 
$\xi_{\pm}(\bm{k})$ is now changed to $\xi_{\pm}(\bm{k})=\pm \varepsilon(\bm{k}) + \frac{a}{2} k^2 -\mu$.

Let us consider the normal state with $\mu>M$. 
There are several contributions to the Meissner kernel. 
First is the contribution from the diamagnetic current which appeared 
in Eq.~(\ref{NewCurrent}) artificially. 
Since this diamagnetic contribution is proportional to the electron density, we obtain
\begin{equation}
K^{\rm dia} = K^{{\rm dia}}_- + K^{{\rm dia}}_+, \qquad 
K^{{\rm dia}}_- = \frac{ae^2}{3\pi^2} k_0^3, \qquad
K^{{\rm dia}}_+ = \frac{ae^2}{3\pi^2} k_F^3,
\label{ArtifKernel1}
\end{equation}
where $k_0$ is the artificial Fermi surface of the lower band shown in Fig. \ref{dis_q}, 
i.e., $\xi_{-}(\bm{k}_0)=0$ is satisfied,
and $k_F$ is the Fermi surface of the upper band.
Sign $\pm$ of the subscript indicates the upper ($\eta=+$) and lower ($\eta=-$) band.
The second contribution is from the intra-band (Fermi-surface) contribution 
from the artificial Fermi surface at $k_0$ (Eq.~(\ref{ArtifKintra}) with $\Delta=0$), 
which is given by
\begin{align}
K^{\mathrm{intra,(N)}}_{xx-} (0,0) &= -\frac{e^2}{3\pi^2} k_0^2 
\frac{(v^2k_0 - ak_0 \varepsilon(\bm{k}_0))^2}{\varepsilon^2(\bm{k}_0)} 
\frac{1}{|ak_0 - \frac{v^2 k_0}{\varepsilon(\bm{k}_0)}|} \notag \\
&= -\frac{e^2}{3\pi^2} k_0^3 \left( a - \frac{v^2}{\varepsilon(\bm{k}_0)} \right).
\label{ArtifKernel2}
\end{align}
Note that $\partial f/\partial E_{-}$ becomes a delta function 
$-\delta(\xi_{-}(\bm{k}))$ when $\Delta=0$, and 
$a - \frac{v^2}{\varepsilon(\bm{k}_0)}>0$ holds since 
$\varepsilon(\bm{k}_0)$ is approximated as 
$\varepsilon(\bm{k}_0)\sim vk_0 \sim 2v^2/a$. 
Similarly, from the Fermi surface at $k_F$, we obtain
\begin{align}
K^{\mathrm{intra,(N)}}_{xx+} (0,0) 
= -\frac{e^2}{3\pi^2} k_F^3 \left( a + \frac{v^2}{\varepsilon(\bm{k}_F)} \right).
\label{ArtifKernel3}
\end{align}
Finally, the inter-band contribution $K^{\mathrm{inter,(N)}}_{xx} (0,0)$ (Eq.~(\ref{ArtifKinter}))
can be calculated analytically. 
The integrand in Eq.~(\ref{ArtifKinter}) is nonzero in the region where $\xi_+(\bm{k})\xi_-(\bm{k})<0$, 
i.e., in the region of $k_F<k(<k_0)$. Thus we obtain
\begin{align}
K^{\mathrm{inter,(N)}}_{xx} (0,0) &= 
\sum_{\bm{k}} F^{\mathrm{inter}}(\bm{k},0) \notag \\
&=  - \frac{e^2}{2\pi^2} \int_{k_F}^{k_0} 
k^2 dk \frac{2v^2}{\varepsilon(\bm{k})} \left( 1- \frac{v^2 k^2}{3\varepsilon(\bm{k})^2} \right) 
\notag \\
&= -\frac{e^2}{3\pi^2} \frac{v^2k^3}{\varepsilon(\bm{k})} \biggr|_{k_F}^{k_0} \notag \\
&= -\frac{e^2}{3\pi^2} \left(
\frac{v^2 k_0^3}{\varepsilon(\bm{k}_0)} - \frac{v^2 k_F^3}{\varepsilon(\bm{k}_F)} \right).
\label{ArtifKernel4}
\end{align}
We can show that the total of Eqs.~(\ref{ArtifKernel1})-(\ref{ArtifKernel4}) vanishes,
which means that the Meissner kernel
in the normal state vanishes.
Next, let us consider the case of superconductivity. 
We assume that the superconducting gap $\Delta(\bk)$ is finite only for $k<k_c<k_0$, as discussed 
in the text. 
In this case, the artificial Fermi surface at $k_0$ survives. 
Thus, the contributions of $K^{{\rm dia}}_-$ and $K^{\mathrm{intra,(N)}}_{xx-} (0,0)$ remain.
On the other hand, $K^{\mathrm{intra,(N)}}_{xx+} (0,0)$ vanishes because of the opening of 
a gap near $k=k_F$. 
Therefore, the total Meissner kernel, $K$, becomes
\begin{align}
K= & K^{\mathrm{inter,(S)}}_{xx} (0,0) + K^{{\rm dia}}_+ + K^{{\rm dia}}_- 
+ K^{\mathrm{intra,(N)}}_{xx-} (0,0)  \notag\\ 
= & K^{\mathrm{inter,(S)}}_{xx} (0,0) +  \frac{e^2}{3\pi^2} \frac{v^2 k_0^3}{\varepsilon(\bm{k}_0)} 
+ \frac{ae^2}{3\pi^2} k_F^3,
\label{ArtifKernelFinal}
\end{align}
where we have used the explicit results in Eqs.~(\ref{ArtifKernel1})-(\ref{ArtifKernel4}).
%Since $a$ is very small, the last term in (\ref{ArtifKernelFinal}) can become arbitrarily small.
%Then the substraction of the normal component of the inter-band contribution 
%$K^{\mathrm{inter,(N)}}_{xx-} (0,0)$ is justified. 
Equation (\ref{ArtifKernelFinal}) means that we should include the second and third terms 
to obtain the correct Meissner kernel.
Since $\Delta(\bk)$ vanishes for $k > k_c$, 
we can rewrite $K_{xx}^{\mathrm{inter,(S)}}(0,0)$ 
[Eq. (\ref{ArtifKinter})] as follows,
\begin{align}
K_{xx}^{\mathrm{inter,(S)}}(0,0) = & \sum_{|\bk| < k_c} F^{\mathrm{inter}}(\bk,\Delta(\bk)) +  \sum_{|\bk| >  k_c} F^{\mathrm{inter}}(\bk,\Delta(\bk)) \notag \\
= & \sum_{|\bk| < k_c} F^{\mathrm{inter}}(\bk,\Delta(\bk)) -\frac{e^2}{3\pi^2} \left[ \frac{v^2k_0^3}{\varepsilon(k_0)} - \frac{v^2k_c^3}{\varepsilon(k_c)} \right],
\end{align}
where we have used the similar calculation 
as in (\ref{ArtifKernel4})
for the integral of $|\bk|>k_c$.
Then, the total Meissner kernel (\ref{ArtifKernelFinal}) can be rewritten as
\begin{equation}
K=\sum_{|\bk|<k_c} F^{\mathrm{inter}}(\bk,\Delta(\bk)) + \frac{e^2}{3\pi^2} \frac{v^2k_c^3}{\varepsilon(k_c)} +\frac{ae^2}{3\pi^2}k_F^3.\label{kernelfin2}
\end{equation}
Since $a$ is a small parameter,
the last term in (\ref{kernelfin2})
can become arbitrarily small.
Furthermore, it is easily seen
that the second term is equal to $-\sum_{|\bk|<k_c}F^{\mathrm{inter}}(\bk,0)$,
i.e., the normal state contribution 
in the region of $0<|\bk|<k_c$.
As a result, in the small $a$ limit,
we obtain 
\begin{equation}
K = \sum_{|\bk|<k_c} F^{\mathrm{inter}}(\bk,\Delta(\bk)) -\sum_{|\bk|<k_c}F^{\mathrm{inter}}(\bk,0) ,
\end{equation}
which is equivalent to Eq. (\ref{true_kernel}).
\par
Although we have used a specific model in this Appendix,
we think that the subtraction of the normal state contribution
as in Eq. (\ref{true_kernel}) is reasonable.

\end{document}